\documentclass[modern]{aastex631}
\submitjournal{ApJ}

\received{July 25, 2024}
\accepted{September 10, 2024}

\shorttitle{FF Galaxy Nucleus}
\shortauthors{Dottori and D\'iaz}

\graphicspath{{./}{figures/}}

\begin{document}

\title{Evidence of a Forming Nucleus in the Fourcade-Figueroa Galaxy}

\author[0000-0002-2000-7141]{H. Dottori}
\affiliation{Instituto de Física, UFRGS, Av. B. Gon\c{c}alves 9500, Porto Alegre, Brasil}
\author[0000-0001-9716-5335]{R.J. Díaz}
\affiliation{Gemini Observatory, NSF NOIRLab, 950 N. Cherry Ave., Tucson, AZ 85719, USA}
\affiliation{Observatorio Astron\'omico de C\'ordoba, UNC, Laprida 854, C\'ordoba, Argentina}






\begin{abstract}

We analyze data from the IRAS, WISE, and Planck satellites, revealing an unresolved dust condensation at the center of the Fourcade-Figueroa galaxy (ESO270-G017), which may correspond to a forming nucleus. We model the condensation's continuum spectrum in the spectral range from 3 to 1300 $\micron$ using the DUSTY code. The best-fit model, based on the Chi-square test, indicates that the condensation is a shell with an outer temperature of $T_\mathrm{out} \approx 12$\,K and an inner boundary temperature of $T_\mathrm{i} \approx 500$\,K. The shell's outer radius is $r_o = 86.2$\,pc, and the inner cavity radius is $r_i = 0.082$\,pc. The condensation produces an extinction $A_V = 50$ mag and its luminosity is $L_c = 1.08 \times 10^{34}$\,W, which would correspond to a burst of massive star formation approximately similar to the central 5\,pc of R\,136 in the LMC and NGC\,3603, the ionizing cluster of a giant Carina arm H{\small{II}} region. The comparison with Normal, Luminous, and Ultra-Luminous Infrared Galaxies leads us to consider this obscured nucleus as the nearest and weakest object of this category.

\end{abstract}


\section{Introduction} 
The discovery of the Fourcade-Figueroa object (ESO\,270-G017, hereinafter FF) was first reported by \citet{1971BAAA...16...10F}, who provided its coordinates and described it as a $4\arcmin$ long 
and $19\arcsec$ wide object without signs of a core. Its extragalactic origin was determined by \citet{1973A&A....23..405D}, who measured a corrected radial velocity of 830\,km\,s$^{-1}$ 
and suggested it might be an irregular galaxy, referred to as a ``shred'' according to \citet{Arp1967ApJ...148..321A}.

\citet{1978PASP...90..237G}, using the CTIO 4\,m telescope, determined a rotational velocity of 10\,km\,s$^{-1}\,(\arcmin)^{-1}$ in the central part of FF, classified it as a late-type spiral galaxy, and estimated its distance to be twice that of the Centaurus Group. He argued that since stars were detected in the 4\,m telescope photographs, FF could not be much further away than the Cen\,A host galaxy (NGC\,5128). In contrast, \citet{1979AJ.....84.1270D} classified FF as SB(s)m and identified it as a background galaxy.

\citet{1992MNRAS.257..689T} simulated a strong prograde encounter between a Milky Way-like progenitor and NGC\,5128 with a mass ratio of 1:10. The collision resulted in FF (the shred), the dwarf elliptical NGC\,5237 (the bulge of the primitive spiral galaxy), and the dust lane around NGC\,5128, which contains about half of the primitive spiral disk and rotates as observed. The model also explained the relative positions and velocities of NGC\,5128, FF, and NGC\,5237, and the hot gas emission from NGC\,5237. However, a modern distance measurement of FF at 6.95\,Mpc by \citet{2015ApJ...805..144K} challenges Thomson's model. Still, the origin of FF remains unknown, due to the difference of 300\,km\,s$^{-1}$ between the FF observed radial velocity and the Hubble flow at 6.95\,Mpc.

\citet{2021A&A...652A.108S} conclude from radio observations that FF has a dark matter halo with a compact core and that the H{\small{I}} disk is considerably larger than the optical one, supporting the disruptive character of this super-thin galaxy, as previously suggested by \citet{1971BAAA...16...10F}.

\begin{figure}[ht]
\centering
\includegraphics[width=0.8\linewidth]{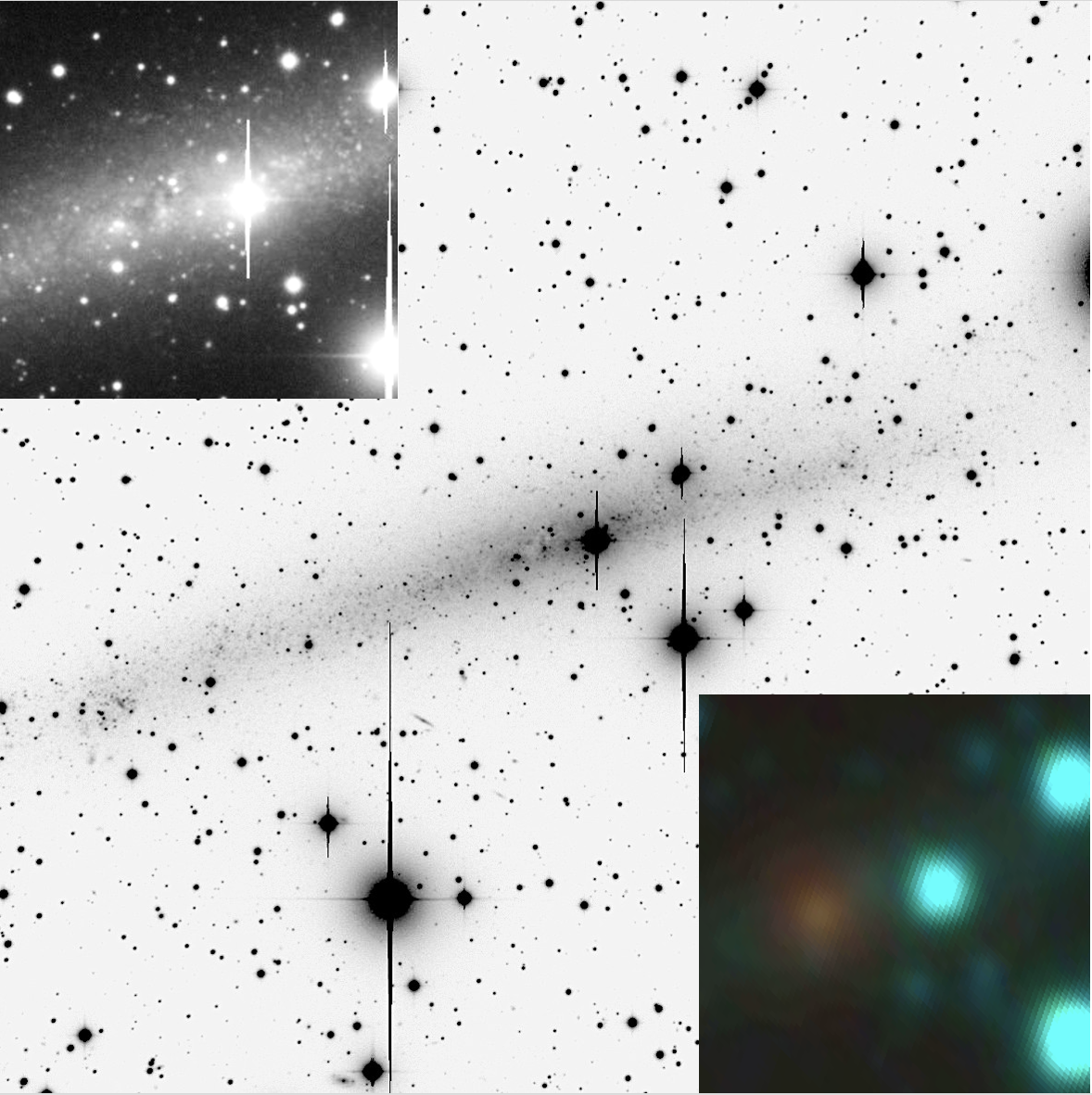}

\caption{
Optical image of the FF Object from the Carnegie-Irvine Atlas of Galaxies, covering an area of $8\arcmin \times 8\arcmin$, composed from images taken using the $B$, $V$, $R$, and $I$ filters. The upper left inset provides a high-contrast zoom into the central $2\arcmin \times 2\arcmin$ area, highlighting some dust patches in the central region. The lower right inset displays a pseudo-color WISE image of the same $2\arcmin \times 2\arcmin$ area, where 3.4\,--\,4.6\,\micron \, light is colored blue, 12\,\micron \,light is green, and 22\,\micron \,light is red.
}

\label{fig:FF1} 
\end{figure}

\citet{1971BAAA...16...10F} claimed that FF does not show a nucleus, and a central concentration is not evident in the optical (DSS) or near-infrared (2MASS) imaging sky surveys. However, we found that WISE satellite images (Fig. \ref{fig:FF1}) reveal an object strongly emitting in the mid-infrared located near the galaxy's isophotal center \citep{1979AJ.....84.1270D}. In Section \ref{sec:obs}, we discuss FF's mid (MIR) and far (FIR) IR archival data. In Section \ref{sec:dusty}, we discuss the parameter space explored with the DUSTY code and the model that best fits FF's IR observations, and in Section \ref{sec:conc}, we discuss FF's IR core properties in comparison to well-known Dust Obscured Galaxies (DOGs) and young star-formation regions.

\begin{figure}[ht]
\centering
\includegraphics[width=0.99\linewidth,height=0.495\columnwidth]{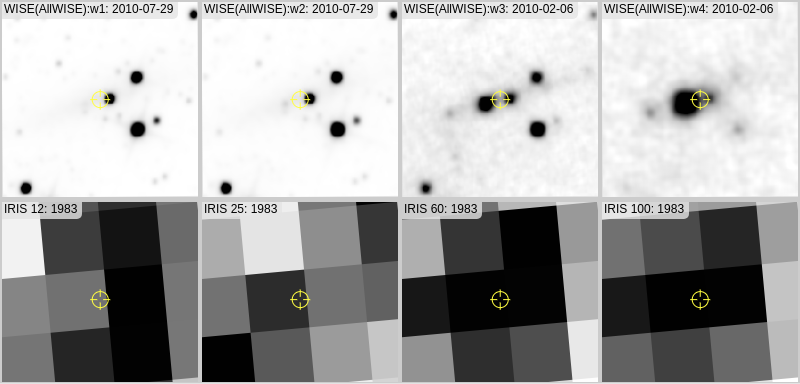}
\caption{
Mosaic of images from WISE and IRIS, IRAS new generation maps, $4\arcmin \times 4\arcmin$ centered on the FF galaxy's isophotal center at RA=13$^h$ 34$^m$ 47$^s$.30, Dec=$-45\arcdeg 32\arcmin 51\farcs$ \citep{1979AJ.....84.1270D}. \citet{2013AJ....145..101K} report the same isophotal center with an uncertainty of $15\farcs0$ in each coordinate.
}

\label{fig:FF2} 
\end{figure}

\begin{figure}[ht]
\centering
\includegraphics[width=0.7\linewidth,height=0.6\columnwidth]{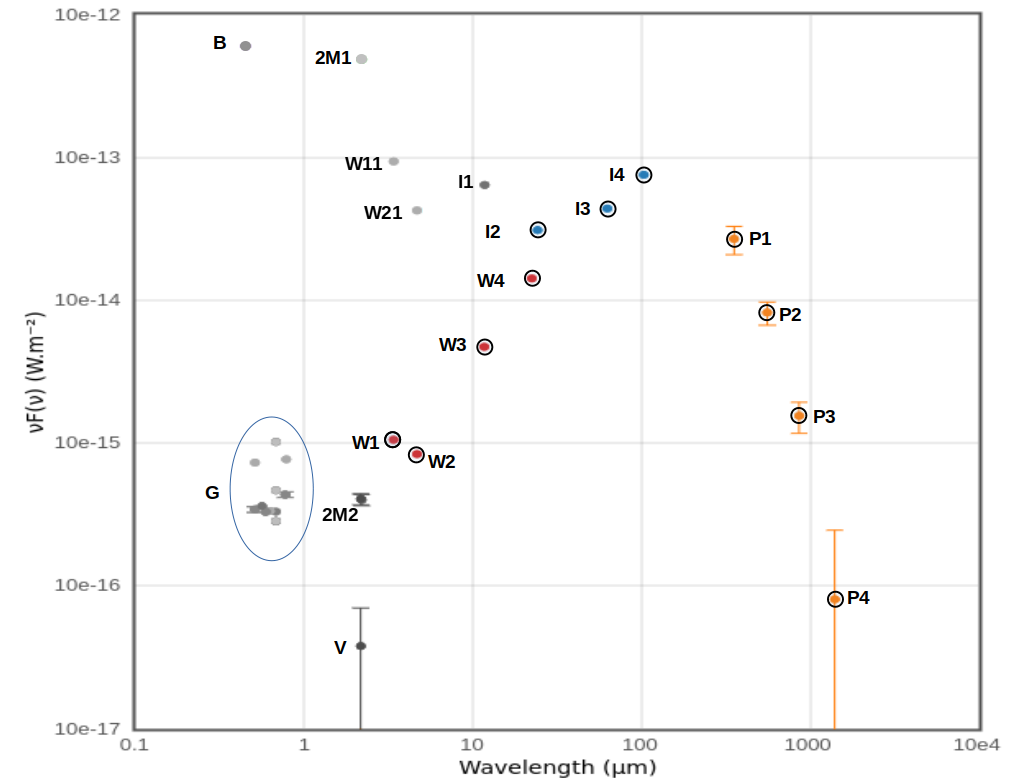}
\caption{
IR and optical sources identified in Table \ref{tab:obs} within a $3\farcs0 \times 3\farcs0$ area centered on the WISE coordinates of the source FFIR. $V$ corresponds to Vista's Ks measurement of a star, and the ellipse includes the brightness of Gaia sources between 0.5 and 0.7 $\mu$m. $B$ and $2M1$ are galaxy integrated brightness in the B and Ks bands.
}
\label{fig:FF3}
\end{figure}

\section{Data Sources} \label{sec:obs}

WISE images (Fig. \ref{fig:FF2}) show a non-stellar, point-like IR source, which we call FFIR, located approximately $18\farcs0$ to the E-SE of FF's isophotal center. The position and brightness of this source are shown in Table \ref{tab:obs}. The first run of WISE observations began on 14 Jan 2010 \citep{Wright2010AJ....140.1868W}; FF was observed on 29 Jul 2010 with detectors cooled to 10\,K. A second run was conducted only at the two shortest wavelengths, $W11$ and $W21$, after the telescope was reactivated on 19 Dec 2013, with the detectors at a temperature of approximately 75\,K \citep{Kourkchi2020ApJ...896....3K}. The brightness difference between $W1$, $W2$ and $W11$, $W21$ is most probably due to the high level of noise in the second round of observations. However, we cannot rule out that there has been an alteration in the brightness of the source between the two observation periods, which could have excited the numerous lines and/or spectral bands in this IR spectral region. In fact, variability in these two filters has recently been reported by \citet{Arevalo2024A&A...683L...8A} for the newly born active nucleus ZTF20aaglfpy. In this study, we only take into account the first epoch observations, as they, together with IRAS and Planck observations, better follow the spectral energy distribution (hereafter SED) that should be expected for cold dust radiation. 

IRAS archive images \citep{Li2011ApJS..197...22L} show continuous emission in the four bands (Fig.~\ref{fig:FF2}), whose centroid is displaced $2\farcs57$ to the E-SE of the WISE source (Table \ref{tab:obs}). As the IRAS spatial resolution ranges from $0\farcm5$ at 12\,\micron\ to $2\farcm0$ at 100\,\micron, we can safely assume that both telescopes are detecting the same source. By comparing $W3$ and $W4$ with $IRAS 12$ and $IRAS 25$, respectively (Table \ref{tab:obs}), we see that the $IRAS 12$ band is likely strongly affected by nearby projected stars from the Milky Way. In contrast, these stars practically do not affect the $IRAS 25$ band. Since stars radiate approximately as black bodies, we assume that $IRAS 60$ and $IRAS 100$ are only measuring FFIR radiation.

Within a $3\farcs0 \times 3\farcs0$ area centered at the WISE coordinates of FFIR, we also find a Planck FIR point-like source \citep{2016A&A...594A..26P} whose position is $1\farcs98$ to the W-NW of the WISE FFIR. Due to the positional uncertainty of Planck and WISE, we also assume that both correspond to the same source. This assumption is justified by the SED shown by the sequence $W1$, $W2$, $W3$, $W4$, $I2$, $I3$, $I4$, $P1$, $P2$, $P3$, and $P4$ (see Fig. \ref{fig:FF3}), which we analyze in Section \ref{sec:dusty}.

Johnson $B$ \citep{2020MNRAS.494.2600N} and 2MASS $2M1$ \citep{2013AJ....145..101K} are the total brightness of the FF galaxy at these bands. 2MASS $2M2$ is a star located $3\farcs45$ to the NE of FFIR, as well as VISTAS $V$ located $2\farcs71$ to the S-SW, which was also observed by GAIA in the optical region. The sources cataloged by Gaia are the points surrounded by the ellipse in Fig.~\ref{fig:FF3} \citep{2018ApJ...867..105T}.

\begin{table}[ht]
\centering
\vspace{0.5cm}
\caption{IR and optical sources within a $3\farcs0 \times 3\farcs0$ area centered on the WISE FFIR source. Caution must be taken when accessing \url{http://vizier.cds.unistra.fr/vizier/sed/}, because there FFIR appears located at the coordinates of the photometric center 
(see Fig. \ref{fig:FF2}).
}
\label{tab:obs}

\begin{tabular}{ccccccc}
\hline\hline
\colhead{Band} & \colhead{AR} & \colhead{DEC} & \colhead{Wavelength} & \colhead{Flux} & \colhead{$\Delta$ Flux} & \colhead{Position} \\
\hline
\colhead{} & \colhead{HH:MM:SS} & \colhead{dd:mm:ss} & \colhead{\micron} & \colhead{W\,m$^{-2}$} & \colhead{W\,m$^{-2}$} & \colhead{Fig.\,2} $^a$\\
\hline

Johnson | B &13:34:49.472 &-45:32:58.54 & 4.44e-1 & 6.06e-13 & & B  \\
2MASS | Ks &13:34:49.472 &-45:32:58.54 & 2.16e+0 & 4.90e-13 & & 2M1 \\
2MASS | Ks &13:34:49.654 &-45:32:48.14 & 2.16e+0 & 4.05e-16 & 3.60e-17 &2M2  \\
Vista | Ks &13:34:49.400 & -45:32:53.65& 2.13e+0 & 3.84e-17 & 3.25e-17 & V  \\ 
WISE | W1 &13:34:49.472 &-45:32:58.54 & 3.35e+0 & 1.06e-15 & 2.68e-17 & W1  \\ 
WISE | W1 &13:34:49.472 &-45:32:58.54 & 3.35e+0 &9.49e-14 & & W12  \\
WISE | W2 &13:34:49.472 &-45:32:58.54 & 4.60e+0 & 8.47e-16 & 1.96e-17 & W2   \\
WISE | W2 &13:34:49.472 &-45:32:58.54 & 4.60e+0 & 4.31e-14 & & W21  \\
WISE | W3 &13:34:49.472 &-45:32:58.54 & 1.16e+1 & 4.77e-15 &7.78e-17 & W3  \\ 
WISE | W4 &13:34:49.472 &-45:32:58.54 & 2.21e+1 & 1.44e-14 & 4.07e-16 & W4  \\ 
IRAS | 12 &13:34:49.688 &-45:32:5.84 & 1.16e+1 & 6.47e-14 & & I1  \\ 
IRAS | 25 &13:34:49.688 &-45:32:59.54 & 2.39e+1 & 3.14e-14 & & I2  \\ 
IRAS | 60 &13:34:49.688 &-45:32:59.54 & 6.18e+1 & 4.44e-14 & & I3   \\
IRAS | 100 &13:34:49.688 &-45:32:59.54 & 1.02e+2 &7.62e-14& & I4  \\ 
Plank | 857GHz &13:34:49.300 &-45:32:57.69 & 3.50e+2 &2.72e-14 &6.08e-15 & P1  \\
Plank | 545GHz &13:34:49.300 &-45:32:57.69 & 5.50e+2 &8.28e-15 &1.53e-15 & P2  \\
Plank | 353GHz &13:34:49.300 &-45:32:57.69 & 8.49e+2 &1.57e-15 & 3.85e-16 & P3  \\
Plank | 217GHz &13:34:49.300 &-45:32:57.69 & 1.38e+3 &8.16e-17 &1.67e-16 & P4  \\
\hline
\end{tabular}
\end{table}
$^a$ {\footnotesize{The last column indicates the source identifications in Fig.\,\ref{fig:FF3}.  $B$: FF total brightness \citep{2020MNRAS.494.2600N}. 
$2M1$: FF total brightness \citep{2013AJ....145..101K}. 
$2M2$: field star. 
$W1$, $W2$, $W3$, and $W4$: observed on 29-07-2010 \citep{2020MNRAS.494.1784A}, \url{http://cdsportal.u-strasbg.fr/?target=ESO270-G017}. 
$W11$ and $W21$: observed after the reactivation on 19-12-2014, \url{https://astro.ucla.edu/~wright/WISE/}. 
I1-I4 IRAS: \url{https://irsa.ipac.caltech.edu/Missions/iras.html}. 
P1-P4: \citep{2016A&A...594A..26P}. }
}

\vspace{0.5cm}
\subsection{Could FFIR be a Background Object?}

In order to check if FFIR might be a background source projected into the central region of the FF galaxy and seen through its inclined disk, we examined all WISE sources within a
$60\farcm0$ radius cone around FFIR (WISE source J133449.44-453257.5). There are 67,475 sources in that area of the sky. For FFIR to be a background source projected onto the galaxy center, 
the putative source would need to be both brighter and bluer than FFIR appears, as its light would have to pass through the dust of the FF disk, whose inclination is 86\arcdeg. We found only seven 
WISE point-like sources that fulfilled these criteria in the considered area. According to \citet{2013AJ....145..101K}, the isophotal center of FF is determined with an uncertainty of $15\farcs0$ 
in each coordinate, and the WISE spatial resolution at the W4 band is $12\farcs0$. These two values led us to adopt a radius of $30\farcs0$ as a reliable elementary cell within which an object projected
onto it could be misinterpreted as a nucleus. These considerations suggest a probability of approximately 1/2000 that FFIR is a background object projected into the central region of FF. Therefore, in what follows, we treat FFIR as a subsystem of the Fourcade-Figueroa galaxy.

\,

\section{Modeling Dust Emission} \label{sec:dusty}
\subsection{Dusty Code} \label{sec:dust}
A quick comparison of FFIR radiated energy with Milky Way dusty clouds excited by individual hot stars \citep{Hirsh2012ApJ...757..113H, Saldano2017RMxAA..53....3S} suggests the presence of a star formation region with tens or hundreds of massive stars.

To solve the problem of radiation transport in a dusty environment, we use the DUSTY 
code \citep{1997MNRAS.287..799I}. The code can handle both spherical and planar geometries. Due to the lack of angular resolution in FFIR observations, we chose to model a dusty spherical shell with a point-like source at the center.

We experimented with a Black Body (BB) source of $T_S = 31,500\,\text{K}$ as in \citet{Hirsh2012ApJ...757..113H} and a source composed of BBs following a \citet{Salpeter1955ApJ...121..161S} IMF, with a highest temperature of 53,000\,K.

DUSTY solves the transfer equation inside the shell by fixing the following parameters:
\begin{enumerate}

\item {\it{Shell chemical composition:}} We use the \citet{Draine1984ApJ...285...89D} proportion of silicates to carbonates (53\% to 47\%) and the inverted relationship as well.

\item {\it{Dust grain sizes:}} We use the standard MSN \citet{Mathis1977ApJ...217..425M} grain size distribution $n(a) \propto a^{-q}$ with $q=3.5$ and $0.005\,\micron \leq a \leq 0.25\,\micron$.

\item {\it{Cloud optical depth at 22\,\micron, $\tau_{22}$:}} We tested 30 values between $1 \leq \tau_{22} \leq 30$.

\item {\it{Density distribution:}} This is treated in terms of the dimensionless profile $\eta(r/r_i)$ where $r_i$ is the radius of the shell's inner boundary. We considered two cases:
    a. Three nested shells with density fall-off softening from $(r/r_i)^{-2}$ to a constant distribution as the radius increases by a factor of 10.
    b. A single shell with a radius $r = 1000 \times r_i$ whose density falls as $(r/r_i)^{-2}$.
    ~$r_i$ scales with the luminosity $L$ as $L^{1/2}$, allowing the models to fit the observations once the temperature of the shell's inner boundary is fixed.

\item {\it{Shell inner boundary temperature, $T_i$:}} This is the only dimensional parameter required to solve the transfer problem. We tested temperatures $200\,\text{K} \leq T_i \leq 1000\,\text{K}$, which are substantially cooler than the dust sublimation temperature (approximately 1500\,K). DUSTY matches the energy released by the central source with $T_i$ by adjusting $r_i$. Inside $r_i$, we have a dust-free cavity.

\end{enumerate}

\subsection{Siebenmorgen and Kr\"ugel 2007 Library.} \label{sec:library}

\citet[S13]{Symeonidis2013MNRAS.431.2317S} used the \citet[SK07]{siebenmorgen2007A&A...461..445S}  library of 7208 radiative transfer models, constrained by five parameters (illustrated by the blue line in their Figure 11 plots), to fit their observations of NIRGs, LIRGs, and ULIRGs. Since SK07 does not provide a dust ball temperature, S13 assigned an ad hoc one by fitting a  single temperature grey-body around the photometric peak in $\nu \times f(\nu)$ (indicated by the green dashed line in their Figure 11 plots). This procedure introduces two additional parameters. S13 notes that, although the SK07 grid offers more flexibility than most standalone SED libraries currently available, it is still too coarse for a complete characterization of the physical properties of high-redshift Herschel samples. Therefore, they chose to use a single parameter to describe the overall shape of the SK07 SED templates: the flux defined as  $\mathcal{F} = \log [L / 4 \pi R^2]$. We cannot model FFIR with SK07 library, as these models do not reach luminosities characteristic of FFIR, and as the authors mention, the SEDs depend on the total luminosity of the source. The faintest objects in the Herschel sample, the NIRGs NGC 253 and M82, are 500 and 1000 times more luminous than FFIR and five times larger (0.359 kpc). Nevertheless, we will compare different scale-free parameters between FFIR and DOGs in the next section.

\begin{figure}[!ht]
\centering
\includegraphics[width=0.65\linewidth,height=0.58\columnwidth]{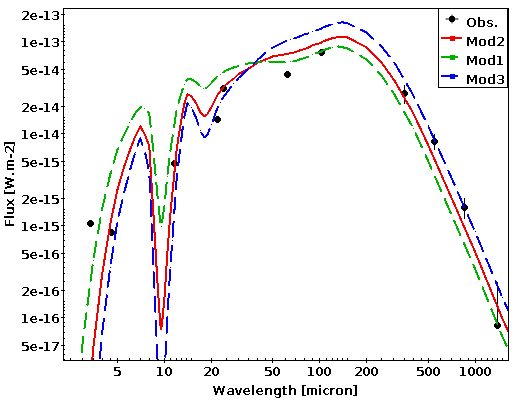}
\caption{The three best DUSTY models compared to Wise, IRAS and Planck's observations.}
\label{fig:FF4}
\end{figure}

\section{Discussion} \label{sec:disc}

We noted that the source, whether an isolated BB or a sum of BBs, does not influence the solution of the radiative transfer equation inside the dust shell. 
DUSTY fits the power of the source to the temperature of the dust shell inner boundary (DSIB) by varying the DSIB size $r_i$ and the ratio $r_i/r_s$, where $r_s$ is the size for the central source to be considered point-like. For the two tested sources, the sum of BBs presents $r_i \approx 1.5$ and $r_i/r_s \approx 4$ times larger than the BB at $T_s = 31,500\,\text{K}$. We will discuss this point later.

The best DSIB temperature is $T_i = 500\,\text{K}$. We tested the \citet{Draine1984ApJ...285...89D} proportion of silicates to carbonates (53\% to 47\%) and the inverse relationship as well. That change influences only the 9.7\,$\micron$ absorption feature, which is deeper for a higher proportion of silicates; however, there are no observations available to differentiate between both metallicities. The models plotted in Fig. \ref{fig:FF4} use the original \citet{Draine1984ApJ...285...89D} metallicity.

A density described by a broken power law gives a better result than a shell with a similar size and density falling as $(r/r_i)^{-2}$ mainly because, at the Planck observation wavelengths, power law models are at least one order of magnitude weaker than the observations.

The optical depth at 22\,$\micron$, $\tau_{22} =$ 2, 3, and 4 furnishes the best models with ${\chi}^2$ values of 9.1, 2.0, and 3.5, respectively. The red line in Fig. \ref{fig:FF4} represents the best model. 

Table \ref{tab:tab2} shows that variations in $\tau_{22}$ do not significantly alter the physical parameters of the models. However, the three SEDs in Fig. \ref{fig:FF4} differ considerably; hence, we conclude that, given a DSBI temperature, the best SEDs are primarily influenced by $\tau(\lambda)$.

The visual extinction is very large, $A_V = 50\,mag$. It lies between that of the NIRG M82 ($A_V = 38\,mag$) and NGC 253 ($A_V = 72\,mag$), as reported by SK07.

The dimensions of the FFIR shell derived from the best parameters are as follows: the DSIB radius is $r_i = 2.7 \times 10^{15}\,\text{m}$ (0.086\,pc), and the shell outer boundary radius is $r_o = 86.2\,\text{pc}$. The FFIR luminosity is $L_{FFIR} = 1.08 \times 10^{34}\,\text{W}$ ($2.8\times10^7 L_\odot$, approximately 20 O3 stars), and the shell outer boundary temperature is $T_o = 11.8 \pm 0.7\,\text{K}$. We used a distance of 6.95\,Mpc \citep{2015ApJ...805..144K}.

\begin{figure}[!ht]
\centering
\includegraphics[width=0.65\linewidth,height=0.58\columnwidth]{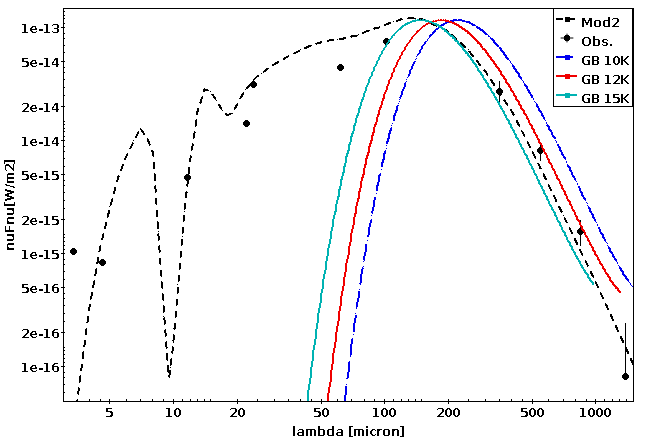}
\caption{The three grey-body models at temperatures of 10\,K, 12\,K, and 15\,K are compared to Model 2 and the observations. The grey-body of 15\,K fits better the emission maximum,  while
the 12\,K one aligns better with the observational data from Planck at wavelengths of 350\,\micron\, and 550\,\micron.
}
\label{fig:FF5}
\end{figure}

\begin{table}[ht]
  \centering
    \begin{tabular}{|c|c|c|c|c|c|c|c|}
        \hline
        ID & $\tau_0$ & $F_1$ (W\,m$^{-2}$) & $r_1$ (m) & $r_1/r_s$ & $\theta_1$ & $T_d$ & $\epsilon $ \\
        \hline
        1 & 2.00E+00 & 1.92E+02 & 3.99E+13 & 4.28E+04 & 2.98E+01 & 12.5 & 2 \\
        2 & 3.00E+00 & 1.89E+02 & 4.02E+13 & 4.32E+04 & 3.00E+01 & 11.8 & 2 \\
        3 & 4.00E+00 & 1.88E+02 & 4.03E+13 & 4.33E+04 & 3.01E+01 & 11.2 & 1 \\
        \hline
    \end{tabular}
    \caption{$\tau_0$, the optical depth at 23\,$\micron$; $F_1$ (W\,m$^{-2}$), the bolometric flux at the inner radius; $r_1$ (m), the inner radius for $L=10^{4}\,L_{\odot}$; $r_1/r_s$, where $r_s$ is the source radius; $\theta_1$, the dust shell inner boundary angular size (in arcseconds) when $F_{\text{bol}} = 10^{-6} \text{ W\,m}^{-2}$; $T_d$, the dust temperature at the outer edge (in K); and $\epsilon$, the maximum error in flux conservation (in \%).
}
    \label{tab:tab2}
\end{table}

The FFIR flux $\mathcal{F} = 9.1$ is not significantly different from that of NIRGs, LIRGs, and ULIRGs, shown in 
S13, their Figure 12. Following the methodology of S13, we fitted a single-temperature grey-body to the photometric peak in $\nu \times f(\nu)$. As shown in Fig. \ref{fig:FF5}, 
a grey-body (GB) at 15\,K best approximates this peak, while the GB at 12\,K fits quite well with the Planck bands at 550 \micron\ and 849 \micron, which are in the same wavelength interval as the Herschel bands fitted by S13.  A comparison of Fig. \ref{fig:FF5} 
with S13's Figure 11 reveals qualitative similarities, with the added benefit that DUSTY also provides the outermost shell surface temperature in a coherent form with other model parameters. We also note that the fit of a GB give a dust ball outermost temperature 20\% to 25\% higher than that obtained from a specific DUSTY model. The FFIR temperature is lower than all temperatures of DOGs in the Herschel sample, as depicted in S13, Figure 15. The color ratios (L110/L250, L70/L100) = (1.6, 0.7) position FFIR in the colder corner of the corresponding color-color diagram, as illustrated in S13, Figure 14, where only a few Herschel sources are located.

\section{Conclusion}\label{sec:conc}

MIR and FIR observations with space telescopes show a point-like dust cloud at the center of the FF galaxy (ESO270-G017), which we have dubbed FFIR. We have modeled the IR SED using the DUSTY code. According to our models, the FF galaxy has experienced a burst of massive star formation (FFIR) similar to the central 5\,pc of R\,136 in the LMC \citep[]{Brands2022A&A...663A..36B, Kalari_2022} and to the ionizing cluster of the giant Carina arm H{\small{II}} region NGC\,3603, whose luminosities are $L_{\mathrm{R136}} = 0.83 \times 10^{34}$\,W and $L_{\mathrm{3603}} = 1.06 \times 10^{34}$\,W \citep{Dreissen1995AJ....110.2235D, Brandl1999Msngr..98...46B, Harayama2008ApJ...675.1319H}, respectively. Since the FFIR massive stars are already enshrouded in their original cocoon, they must be much younger than their counterparts in the LMC and the Milky Way, probably not older than a few hundred thousand years. The FFIR might not be a cluster of stars enshrouded in a single cocoon but a sum of individual cocoons; nevertheless, \citet{JohnsonK2005IAUS..227..413J} pointed out that the Milky Way newborn star clusters' properties appear similar to those of ultracompact H{\small{II}} regions but scaled up in total mass and luminosity, indicating that the model adopted for FFIR would be a good approximation.

A caveat of our solution is that DUSTY requires the exciting source to be point-like. As Table \ref{tab:tab2} shows, for that to happen, the source must have a radius approximately forty thousand times smaller than the DSIB, putting the source size on the order of 1.0\,AU, which is too small when compared to what is known about star formation in the nearby universe. For example, the Arches cluster near the center of our galaxy \citep{Cotera1996ApJ...461..750C} presents $2\times 10^4\,M_{\odot}$ within 0.4\,pc, with a central mass density of $2\times 10^5 M_{\odot}\,\text{pc}^{-3}$, making it the densest known in the Milky Way \citep{Espinosa2009A&A...501..563E}. Bursts of star formation in the nuclear region of galaxies do not necessarily entail large concentrations of stars, as seen in the prototype star-forming galaxies NGC\,7714 \citep{Gonzales1999ApJ...513..707G} and NGC\,604 \citep{Hunter1996ApJ...456..174H}. A possible solution to this dilemma could be microquasars, which are compatible with FFIR from an energetic point of view, as in the cases of S\,236 in NGC\,7793 \citep{2010AIPC.1248..127S}, S\,2 in M\,83 \citep{Soria_2020}, and SS\,433 in the Milky Way \citep{Blundell2004ApJ...616L.159B}. However, microquasars present a Riley-Fanaroff\,2 morphology and mainly deliver mechanical energy, at least in the few known cases and the stage of evolution they are detected today. Another promising solution could be a soft proto-AGN, which we know must exist, as the transformation of a normal nucleus of a galaxy into an active one has been recently reported by \citet{Arevalo2024A&A...683L...8A}.

The possible brightness difference between $W1$, $W2$ and $W11$, $W21$ might be an important issue that deserves a more detailed study, as well as the difference of 300\,km\,s$^{-1}$ between the FF observed velocity and the Hubble flow at 6.95\,Mpc. 

The comparison with DOGs demonstrates that FFIR's properties align well within this category, suggesting that FFIR may be characterized as the weakest DOG ever detected.

The issues highlighted suggest that further observations in the MIR, FIR, and radio-interferometry are crucial to illuminate the nature of FFIR, an object in the local universe that seems to be in the early stages of its evolution since formation.

\begin{acknowledgments}

    This paper has been partially supported with a grant from CNPq of Brazil. This research has made use of the VizieR catalogue access tool, CDS, Strasbourg, France. 
    \software {We processed figures with TOPCAT \citep{Taylor2005ASPC..347...29T}. We used ChatGPT  (https://openai.com/blog/chatgpt) to review the LaTeX symbols and English grammar, a task we previously carried out with human revisors. ChatGPT was also used to generate Python scripts that adjusted DUSTY models and gray-bodies to the observations using Chi-square tests.}

\end{acknowledgments}

\bibliography{FF.bib}{}
\bibliographystyle{aasjournal}

\end{document}